\documentclass[3p,times,twocolumn]{elsarticle}

\usepackage{ecrc,subfigure,}

\volume{00}

\firstpage{1}

\journalname{Nuclear and Particle Physics Proceedings}

\runauth{}

\jid{nppp}

\jnltitlelogo{Nuclear and Particle Physics Proceedings}

\usepackage{amssymb}

\usepackage[figuresright]{rotating}

\newcommand{\AddrHEPHY}{%
 \it Institut f\"ur Hochenergiephysik der \"Osterreichischen Akademie
der Wissenschaften, A-1050 Vienna, Austria\\}

\newcommand{\AddrVienna}{
\it Universit\"at Wien, Fakult\"at f\"ur Physik,
A-1090 Vienna, Austria \\}

\newcommand{\AddrGAKUGEI}{%
 \it Department of Physics, Tokyo Gakugei University, Koganei,
Tokyo 184-8501, Japan\\}

\def\d              {\delta}
\def\ti              {\tilde}
\def\su                {\ti{u}}
\def\dll            {\d^{LL}_{23}}
\def\durr            {\d^{uRR}_{23}}
\def\durl            {\d^{uRL}_{23}}
\def\dulr            {\d^{uLR}_{23}}
\def\x               {\chi}
\def\gev             {{\rm GeV}}

\def\su                {\ti{u}}
\def\sd                {\ti{d}}
\def\sg              {\ti g}
\newcommand{\nn}{\nonumber\\}
\def\bea            {\begin{eqnarray}}
\def\eea            {\end{eqnarray}}
\def\a              {\alpha}

\newcommand{\mch}[1] {m_{\ti \x^+_{#1}}}
\newcommand{\mnt}[1] {m_{\ti \x^0_{#1}}}

\newcommand{\msg}    {m_{\ti g}}
\newcommand{\msu}[1] {m_{\ti u_{#1}}}
\newcommand{\msd}[1] {m_{\ti d_{#1}}}
\def \sca                 {\ti{c}}
\def\st                 {\ti{t}}

\begin{document}

\begin{frontmatter}




\title{Quark-flavour violating Higgs decays to charm and bottom pairs in the MSSM}


\author[a]{E. Ginina}
\author[a]{H. Eberl}
\author[b]{A. Bartl}
\author[c]{K. Hidaka}
\author[a]{W. Majerotto}

\address[a]{\AddrHEPHY}
\address[b]{\AddrVienna}
\address[c]{\AddrGAKUGEI}

\begin{abstract}
We calculate the decay width of $h^0 \to b \bar{b}$ in the Minimal Supersymmetric
Standard Model (MSSM) with quark-flavour violation (QFV) at full one-loop
level. The effect of $\tilde{c}-\tilde{t}$ mixing and $\tilde{s}-\tilde{b}$ mixing is studied
taking into account the constraints from the B-meson data. We discuss and
compare in detail the decays $h^0 \to c \bar{c}$ and $h^0 \to b \bar{b}$ within the framework
of the perturbative mass insertion technique using the Flavour Expansion
Theorem. The deviation of both decay widths from the Standard Model results can
be quite large. While in $h^0 \to c \bar{c}$ it is almost entirely due to the
flavour violating part of the MSSM, in $h^0 \to b \bar{b}$ it is mainly due to the
flavour conserving part. Nevertheless, $\Gamma(h^0 \to b \bar{b})$ can fluctuate up to $\sim 7\%$
due to QFV chargino exchange with large $\tilde{c}-\tilde{t}$ mixing. 
\end{abstract}


\end{frontmatter}

\section{Introduction}
\label{sec:intro}

So far, the Higgs boson properties measured at the LHC experiments are consistent with the Standard Model (SM) 
predictions. Deviations are, however, not yet excluded and could indicate physics beyond the SM. For instance, in the Minimal Supersymmetric Standard Model (MSSM), the discovered Higgs boson can well be the lightest neutral Higgs boson, $h^0$, with a mass of 125 GeV and SM-like couplings. Non-minimal quark-flavour violation (QFV) in the squark sector of the MSSM can additionally affect the Higgs interactions at one-loop level, still being consistent with the B-physics constrains. We study two important decays of the Higgs boson: into a pair of bottom quarks and into a pair of charm quarks at one-loop level in the MSSM with general squark mixing. We consider mainly mixing between the two heavy generations up- and down-squarks, i.e. $\tilde{c}_{\rm L,R}-\tilde{t}_{\rm L,R}$ and $\tilde{s}_{\rm L,R}-\tilde{b}_{\rm L,R}$mixing. We investigate numerically the influence of such mixing on the Higgs properties, taking into account the constraints on quark-flavour mixing from B-physics. This talk is based on Ref.~\cite{Eberl:2016aox}, for more details see the original publication.

\section{QFV in the squark sector of the MSSM}
\label{sec:sq.matrix}

We define the QFV parameters in the up-type squark sector of the MSSM
as follows:
\begin{eqnarray}
\delta^{LL}_{\alpha\beta} & \equiv & M^2_{Q \alpha\beta} / \sqrt{M^2_{Q \alpha\alpha} M^2_{Q \beta\beta}}~,
\label{eq:InsLL}\\[3mm]
\delta^{uRR}_{\alpha\beta} &\equiv& M^2_{U \alpha\beta} / \sqrt{M^2_{U \alpha\alpha} M^2_{U \beta\beta}}~,
\label{eq:InsRR}\\[3mm]
\delta^{uRL}_{\alpha\beta} &\equiv& (v_2/\sqrt{2} ) T_{U\alpha \beta} / \sqrt{M^2_{U \alpha\alpha} M^2_{Q \beta\beta}}~,
\label{eq:InsRL}
\end{eqnarray}
where $\alpha,\beta=1,2,3 ~(\alpha \ne \beta)$ denote the quark flavours $u,c,t$, and $v_{2}=\sqrt{2} \left\langle H^0_{2} \right\rangle$.
$M_{Q,U}$ are the hermitian soft SUSY-breaking squark mass matrices and $T_{U}$ are the soft SUSY-breaking trilinear 
coupling matrices of the up-type squarks. These parameters enter the left-left, right-right and left-right blocks of the $6\times6$ up-type squark mass matrix in the super-CKM basis~\cite{Allanach:2008qq},
\begin{equation}
    {\cal M}^2_{\tilde{u}} = \left( \begin{array}{cc}
        {\cal M}^2_{\tilde{u},LL} & {\cal M}^2_{\tilde{u},LR} \\[2mm]
        {\cal M}^2_{\tilde{u},RL} & {\cal M}^2_{\tilde{u},RR} \end{array} \right)\,.
 \label{EqMassMatrix1}
\end{equation}
The different blocks in eq.~(\ref{EqMassMatrix1}) are given by
\begin{eqnarray}
    & &{\cal M}^2_{\tilde{u},LL} = V_{\rm CKM} M_Q^2 V_{\rm CKM}^{\dag} + D_{\tilde{u},LL}{\bf 1} + \hat{m}^2_u, \nonumber \\
    & &{\cal M}^2_{\tilde{u},RR} = M_U^2 + D_{\tilde{u},RR}{\bf 1} + \hat{m}^2_u, \nonumber \\
    & & {\cal M}^2_{\tilde{u},RL} = {\cal M}^{2\dag}_{\tilde{u},LR} =
    \frac{v_2}{\sqrt{2}} T_U - \mu^* \hat{m}_u\cot\beta\,,
     \label{RLblocks}
\end{eqnarray}
where
$\mu$ is the higgsino mass parameter, $\tan\beta$ is the ratio of the vacuum expectation values of the neutral Higgs fields $v_2/v_1$, with $v_{1,2}=\sqrt{2} \left\langle H^0_{1,2} \right\rangle$, and $\hat{m}_{u}$ is the diagonal mass matrix of the up-type quarks.
Furthermore, $D_{\tilde{u},LL} = \cos 2\beta\, m_Z^2 (T_3^u-e_u
\sin^2\theta_W)$ and $D_{\tilde{u},RR} = e_u \sin^2\theta_W \times$ $ \cos 2\beta \,m_Z^2$, with
$T_3^u$ and $e_u$ being the isospin and
electric charge of the up-type quarks (squarks), respectively, and $\theta_W$ is the weak mixing
angle. $V_{\rm CKM}$ is the Cabibbo-Kobayashi-Maskawa matrix, which we approximate with the unitary matrix.
The up-squark mass matrix is diagonalized by the $6\times6$ matrices $U^{\tilde{u}}$, such that
\begin{eqnarray}
&&U^{\tilde{u}} {\cal M}^2_{\tilde{u}} (U^{\tilde{u} })^{\dag} = {\rm diag}(m_{\tilde{u}_1}^2,\dots,m_{\tilde{u}_6}^2)\,,
\label{Umatr}
\end{eqnarray}
with the mass hierarchy $m_{\tilde{u}_1} < \dots < m_{\tilde{u}_6}$.
The physical mass eigenstates
$\su_i, i=1,...,6$ are given by $\su_i =  U^{\su}_{i \alpha} \su_{0\alpha} $. The QFV parameters of the down squark sector are defined analogously, see Ref.~\cite{Eberl:2016aox}.

We mainly focus on the $\ti{c}_R - \ti{t}_L$, $\ti{c}_L - \ti{t}_R$, $\ti{c}_R - \ti{t}_R$, and $\ti{c}_L - \ti{t}_L$ mixing, 
which is described by the QFV parameters $\delta^{uRL}_{23}$, 
$\delta^{uLR}_{23} \equiv ( \delta^{uRL}_{32})^*$, $\delta^{uRR}_{23}$, and $\dll$, respectively. The
$\ti{t}_R - \ti{t}_L$ mixing is described by the quark-flavour conserving (QFC) parameter $\delta^{uRL}_{33}$.  All the QFV and QFC parameters are assumed to be real.

\section{The processes}
\label{sec:processes}

The decay width of $h^0 \to q \bar{q}$, with $q=c,b$, including one-loop contributions, can be written as
\begin{equation}
\hspace*{-0.5cm}
\Gamma(h^0 \to q \bar{q})=\Gamma^{\rm tree}(h^0 \to q \bar{q})+\delta \Gamma^{\rm  1loop}(h^0 \to q \bar{q})
\label{decaywidth}
\end{equation}
with the tree-level decay width
\begin{equation}
\Gamma^{\rm tree}(h^0 \to q \bar{q})=\frac{\rm N_C}{8 \pi} m_{h^0} (s_1^q)^2 \bigg( 1- \frac{4 m_q^2}{m_{h^0}^2}\bigg)^{3/2}\,,
\label{decaywidttree}
\end{equation}
where ${N_C}=3$, $m_{h^0}$ is the on-shell mass of $h^0$ and the tree-level coupling $s_1^q$ is given by
\begin{equation}
\hspace*{-0.5cm}
 s_1^{c,b}=\mp g \frac{m_{c,b}}{2 m_W} \frac{\cos{\alpha}}{\sin{\beta}}\left(\frac{\sin{\alpha}}{\cos{\beta}}\right) =\mp\frac{h_{c,b}}{\sqrt{2}}\cos{\alpha}\,(\sin{\alpha})\,,
 \label{treecoup}
\end{equation}
$\alpha$ is the mixing angle of the two CP-even Higgs bosons, $h^0$ and $H^0$.
%
\subsection{Gluino contribution to $h^0 \to c \bar c$}
\label{SecMI:gluino2ccb}

\begin{figure*}[ht!]
\tiny{
\centering
\subfigure[]{
   { \mbox{\hspace*{-1cm} \resizebox{4.8cm}{!}{\includegraphics{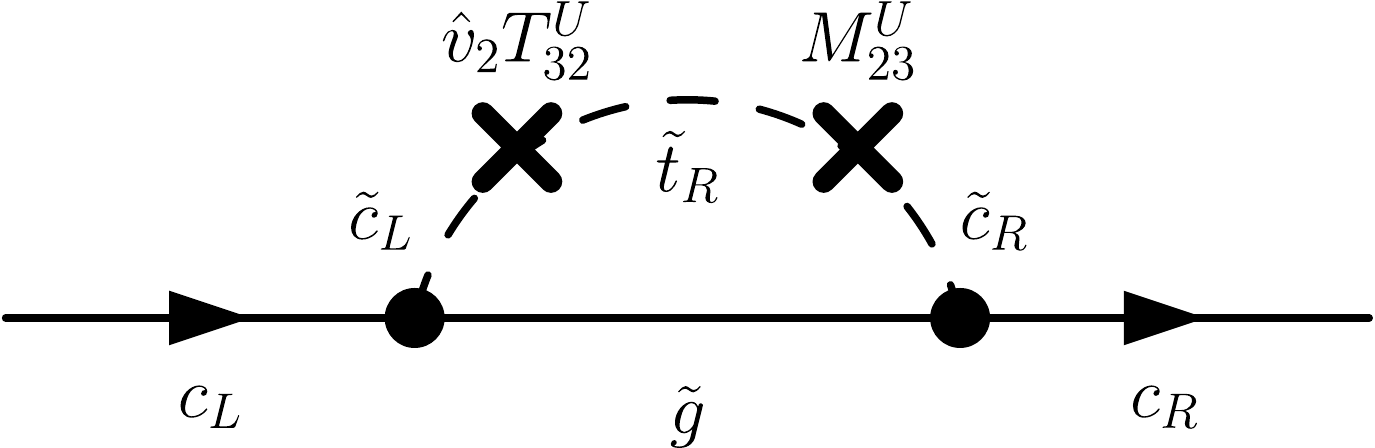}} \hspace*{-0.8cm}}}
   \label{diag_ccb1}} \qquad
 \subfigure[]{
   { \mbox{\hspace*{+0.cm} \resizebox{4.8cm}{!}{\includegraphics{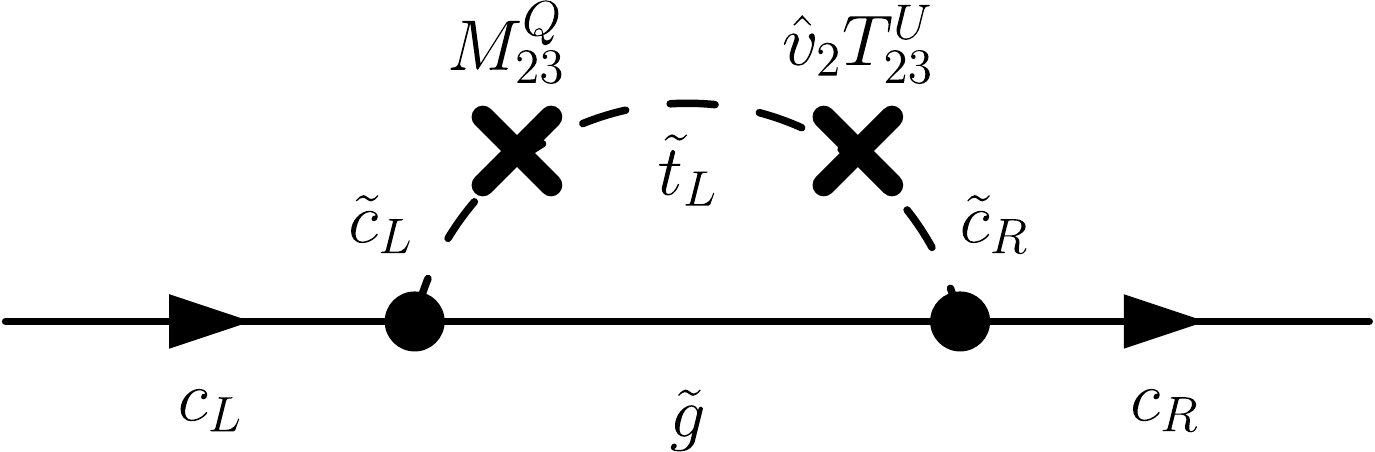}} \hspace*{-1cm}}}
  \label{diag_ccb2}
  }\\
 \subfigure[]{
   { \mbox{\hspace*{-1cm} \resizebox{4.8cm}{!}{\includegraphics{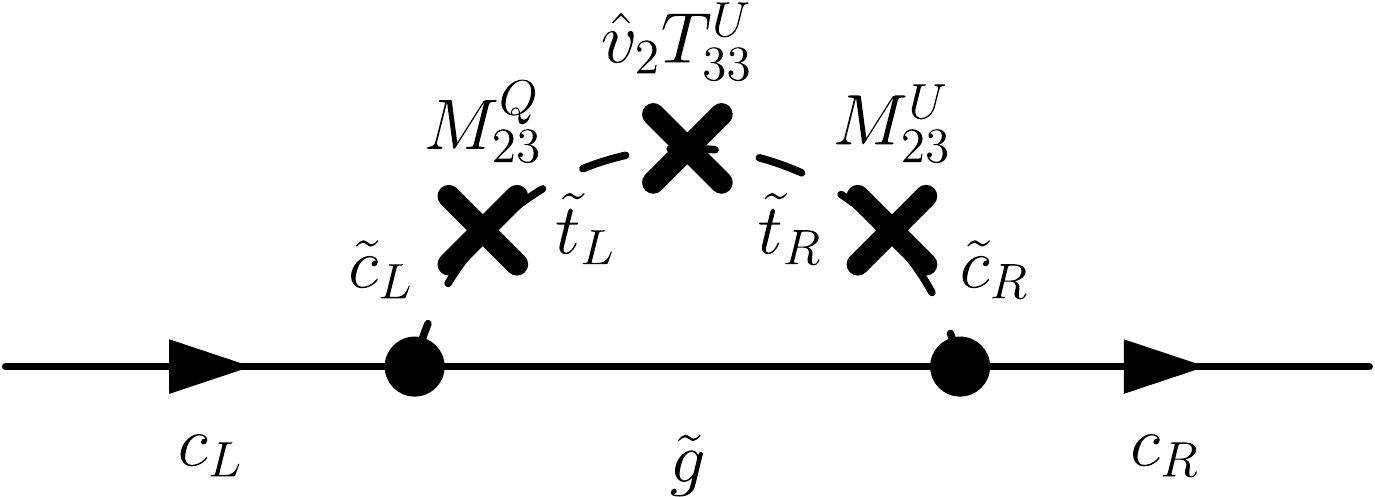}} \hspace*{-0.8cm}}}
   \label{diag_ccb3}} \qquad
 \subfigure[]{
   { \mbox{\hspace*{+0.cm} \resizebox{4.8cm}{!}{\includegraphics{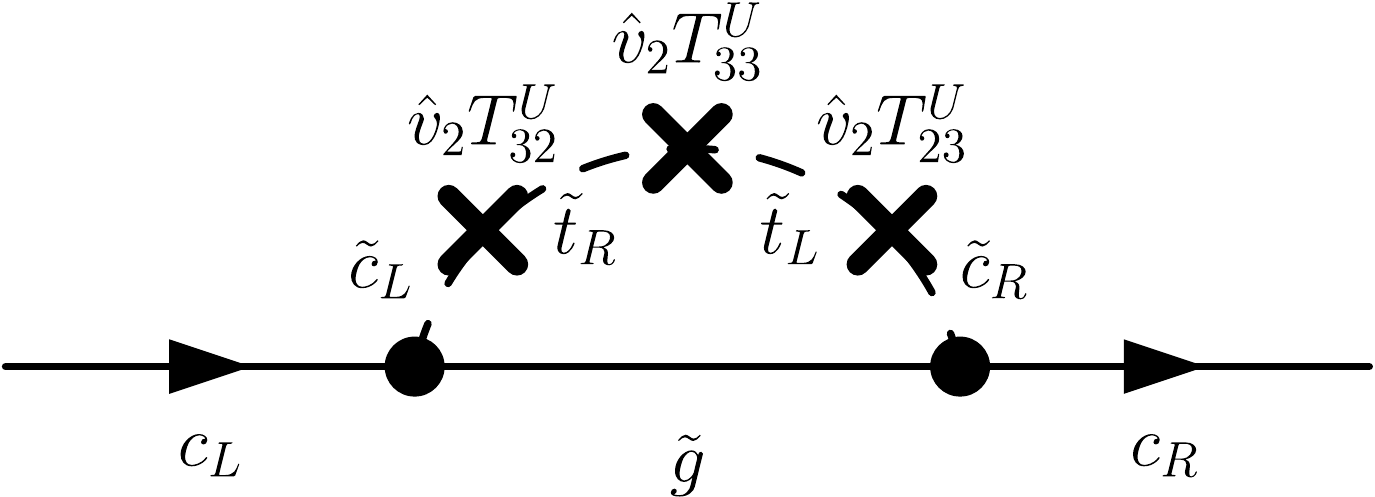}} \hspace*{-1cm}}}
  \label{diag_ccb4}
  } \caption{Quark-flavour violating mass insertions to the charm quark self-energy with gluino, corresponding to
  $T_2$ and $T_3$ in eq.~(\ref{MI_Ts}).}}
\label{diag_ccb}
\end{figure*}
For the calculation of $\delta \Gamma^{\rm  1loop}(h^0 \to b \bar{b})$ we proceed in a way analogous to the calculation of $\delta \Gamma^{\rm  1loop}(h^0 \to c \bar{c})$ in Ref.~\cite{Bartl:2014bka}.
The corresponding loop diagrams are shown in Fig.~2 of \cite{Bartl:2014bka}, with the replacements: $c \leftrightarrow b$ and $\tilde{u} \leftrightarrow \tilde{d}$. The dominant supersymmetric (SUSY) contribution is due to gluino and chargino exchange, which also contribute to the self-energy of the b-quark.

As in Ref.~\cite{Bartl:2014bka}, in our calculation we employ the $\overline{\rm DR}$ renormalisation scheme, with the Lagrangian input parameters, defined at the scale $Q=1~\rm TeV$. The shifts from the $\overline{\rm DR}$ masses and fields to the physical scale-independent quantities are obtained using on-shell renormalisation conditions. To assure infrared (IR) convergence we include the real gluon/photon radiation contributions as well.

Furthermore, we compare and recalculate in the mass insertion (MI) technique $\Gamma(h^0 \to b \bar{b})$ and $\Gamma(h^0 \to c \bar{c})$, as previously
studied in Ref.~\cite{Bartl:2014bka}.
In this context we often refer to the one-loop representation
\begin{equation}
\Gamma(h^0 \to b \bar{b}) = \Gamma^{g, \rm impr} + \d \Gamma ^{\tilde{g}} + \d \Gamma ^{EW}\,,
\end{equation}
where $ \Gamma^{g, \rm impr}$ includes the tree-level and the gluon one-loop contribution (see eq.(55) in~\cite{Bartl:2014bka}), $\delta \Gamma ^{\tilde{g}}$  is the gluino one-loop contribution, and $ \delta \Gamma ^{EW}$ is the electroweak one-loop contribution.

\section{Mass insertion technique}
\label{sec:MI}

The perturbative interaction between the Higgs and the squarks is explicitly proportional to the soft SUSY-breaking trilinear coupling matrices, $T_{U,D}$. However, the dependence on the soft SUSY-breaking mass matrices, ${\cal M}^2_{\tilde{u},\sd}$ is hidden in the squark mixing matrices, $U^{\tilde{u}, \tilde{d}}$, which makes the analysis complicated. An effective approach using the mass insertion (MI) approximation gives access to the explicit dependences on these QFV parameters and allows an analytic approach to study the QFV effects. In our calculations we exploit the Flavour Expansion Theorem (FET)~\cite{Dedes:2015twa}.   

In the following we briefly review the main suggestion of the MI approximation. If $X = U^{\tilde q}_{i A} U^{\tilde q *}_{i B} B_0(0, m^2, m^2_{\tilde q_i})$, where $B_0$ is the two-point function given in terms of mass eigenstates, and $U^{\tilde q}$ are the rotation matrices defined with eq.~(\ref{Umatr}) ($A \neq B$), then $X$ 
can be expanded into mass insertions (MIs) by the FET~\cite{Dedes:2015twa}:
\begin{eqnarray}
\hspace*{-2cm}
X & =  & M^I_{A B} \, b_0\! \left(1,  m^2, \{M_{A A}, M_{B B}\}\right) \nonumber \\
   & + &   M^I_{A i} M^I_{i B} \, b_0\! \left(2,  m^2, \{M_{A A}, M_{i i},  M_{B B}\} \right) \nonumber \\
   & + &   M^I_{A i} M^I_{i j} M^I_{j B} \, b_0\! \left(3,  m^2, \{M_{A A}, M_{i i}, M_{j j}, M_{B B}\} \right) \nonumber \\
   & + &   M^I_{A i} M^I_{i j} M^I_{j k}  M^I_{k B} \, b_0\, \big(4,  m^2,  \nonumber \\
   && \{M_{A A}, M_{i i}, M_{j j}, M_{k k},  M_{B B}\} \big) + \ldots \, ,
\label{mass2MI_exp4}   
\end{eqnarray}
The insertions are repesented by the elements of the matrix $M^I$, with $M^I_{ii}=0$. 
The generalized $b_0$ functions used in eq.~(\ref{mass2MI_exp4}), where the first argument shows the number of insertions, can be written recursively as~\cite{Dedes:2015twa}
\begin{eqnarray}
\hspace*{-1cm}
 && b_0(1,a,\{b,c\})  =  \frac{b_0(a,b)-b_0(a,c)}{b-c}\,, \nonumber \\
 && b_0(2,a,\{b,c,d\})  = \nonumber \\
 &&   \hspace*{1cm}  \frac{b_0(1,a,\{b,c\})  - b_0(1,a,\{b,d\}) }{c-d}\,, \nonumber \\
 && b_0(3,a,\{b,c,d,e\}) = \nonumber \\
 &&  \hspace*{1cm}   \frac{b_0(2,a,\{b,c, d\})  - b_0(2,a,\{b,c, e\}) }{d-e}\,, \nonumber \\
 && b_0(4,a,\{b,c,d,e,f\}) = \nonumber \\
 &&   \frac{b_0(3,a,\{b,c,d,e\})  - b_0(3,a,\{b,c,d,f\}) }{e-f}\,,
\end{eqnarray}
where
\bea
&& b_0(a,b) \equiv  B_0(0, a,b) = \nn
&&  \hspace*{0.5cm}  = \frac{b \log \left(\frac{b}{Q^2}\right)-a \log
   \left(\frac{a}{Q^2}\right)}{a-b}+ \Delta +1\,,
\eea
with the renormalisation scale $Q$ and the UV-divergence parameter $\Delta$.

\subsection{Gluino contribution to $h^0 \to c \bar c$}
\label{SecMI:gluino2ccb}
In order to demonstrate how the mass insertion approximation works we calculate the self-energy 
of the c-quark $\Sigma_c$ with $\sg$ and $\su_i$ in the loop.
The relevant $LR$-part $\Sigma^{LR}_c = \Sigma^{RL}_c$ reads
\begin{equation}
\Sigma^{LR, \tilde g}_c =   - \frac{2 \a_s}{3 \pi } \frac{\msg}{m_{c}} \sum_{i=1}^6 U^{\su *}_{i2}U^{\su}_{i5}
B_0(m_c^2, m^2_{ \tilde g}, m^2_{\tilde u_i})\, ,
\label {dsigLRsg_c}
\end{equation}
where we assume that the squared $\tilde u$-mass matrix ${\cal M}^2_{\tilde{u}} \equiv M_{ij}$ (see eq. (\ref{EqMassMatrix1})) is in the form
\begin{equation}
\hspace*{-0.7cm}   \small{ {\cal M}^2_{\tilde{u}}=       
         \left( \begin{array}{cccccc}
         M^{LL}_{11} & 0 & 0 & 0 & 0 & 0 \\
         0  & M^{LL}_{22} & M^Q_{23} & 0 & 0 & \hat v_2 T^U_{32} \\
         0 & M^Q_{23} & M^{LL}_{33} & 0 &  \hat v_2 T^U_{23} & \hat v_2 T^U_{33}\\
         0 & 0 & 0 &  M^{RR}_{11} & 0 & 0\\
         0 & 0 &   \hat v_2 T^U_{23} & 0 & M^{RR}_{22} &  M^U_{23} \\
         0 & \hat v_2 T^U_{32} & \hat v_2 T^U_{33} & 0 &  M^U_{23} & M^{RR}_{33}
         \end{array}\right)},     
 \label{EqMassMatrix}
\end{equation}
with $\hat v_2 = v \sin\beta/\sqrt{2} \sim $ 170~GeV, and the QFV elements of the  $3 \times 3$ matrices $M^2_Q$ and $M^2_U$ are denoted by $M^Q_{ij}$ and $M^U_{ij}$, respectively. Using the FET (eq.~(\ref{mass2MI_exp4})) we get 
\begin{equation}
m_c \Sigma^{LR, \tilde g}_c =   - \frac{2 \a_s}{3 \pi } \msg (T_2 + T_3 + T_4 + \ldots)\,,
\label {dsigLRsg_c1}
\end{equation}
where the QFV contributions read
\begin{eqnarray}
T_2 & = & \hat v_2 T^U_{32} M^U_{23}  \, b_0\! \left(2,  \msg^2, \{ M^{LL}_{22}, M^{RR}_{33}, M^{RR}_{22}\} \right) \nn
&+&          \hat v_2 T^U_{23} M^Q_{23}  \, b_0\! \left(2,  \msg^2, \{M^{LL}_{22}, M^{LL}_{33}, M^{RR}_{22}\} \right) \nn
T_3 & = &  \hat v_2 T^U_{33} \left( M^Q_{23}  M^U_{23}  +  3^\rho\, {\hat v_2}^2  T^U_{23}  T^U_{32} \right)\nn
   && \times~b_0\! \left(3,  \msg^2, \{ M^{LL}_{22}, M^{LL}_{33}, M^{RR}_{33}, M^{RR}_{22}\} \right)\, , 
\label{MI_Ts}
\end{eqnarray}
with $\rho = 0$. The corresponding to the terms $T_2$ and $T_3$ graphs  are shown in Figs.~1(a) and 1(b) or 1(c) and 1(d), respectively.
Note that there is no contribution with no mass insertion because of the helicity flip, and also practically no contribution with 
only one insertion, because $T^U_{22}\approx 0$. 
Thus, all terms in eq.~(\ref{MI_Ts}) are quark-flavour violating.

The vertex contributions with $\sg$, $\su^*_i$ and $\su_j$ in the loop, defined by ${\cal L} = - h^0 \bar c\, (c^v_L P_L + c^v_R P_R)\, c$, are calculated in an analogous way. Here we only show the result for the mass insertion expansions for the coefficients $c^v_L$ and $c^v_R$, for details see~Ref.\cite{Eberl:2016aox}. In case of real input parameters, $c^v_L = c^v_R = c^v$, we obtain
\begin{equation}
c^v=  - \frac{2 \a_s}{3 \pi } \msg  {\cos\alpha \over \sqrt2} (T^v_1 + T^v_2 + \ldots)\, ,
\label{coeff_verth02ccb_MI}
\end{equation}
where
\begin{eqnarray}
T^v_1 & = & T^U_{32} M^U_{23}  \, b_0\! \left(2,  \msg^2, \{ M^{LL}_{22}, M^{RR}_{33}, M^{RR}_{22}\} \right)  \nn
                   &+&T^U_{23} M^Q_{23}  \, b_0\! \left(2,  \msg^2, \{M^{LL}_{22}, M^{LL}_{33}, M^{RR}_{22}\} \right)\,, \nn
T^v_2 & = & T^U_{33} \left(M^Q_{23}  M^U_{23} + 3\, {\hat v_2}^2 \, T^U_{23}  T^U_{32} \right) \nn
&& \times~ b_0\! \left(3,  \msg^2, \{ M^{LL}_{22}, M^{LL}_{33}, M^{RR}_{33}, M^{RR}_{22}\} \right) \,. 
\label{MI_Tvs}
\end{eqnarray}
\begin{figure*}[ht]
\centering
\subfigure[]{
   { \mbox{\hspace*{-1cm} \resizebox{7.cm}{!}{\includegraphics{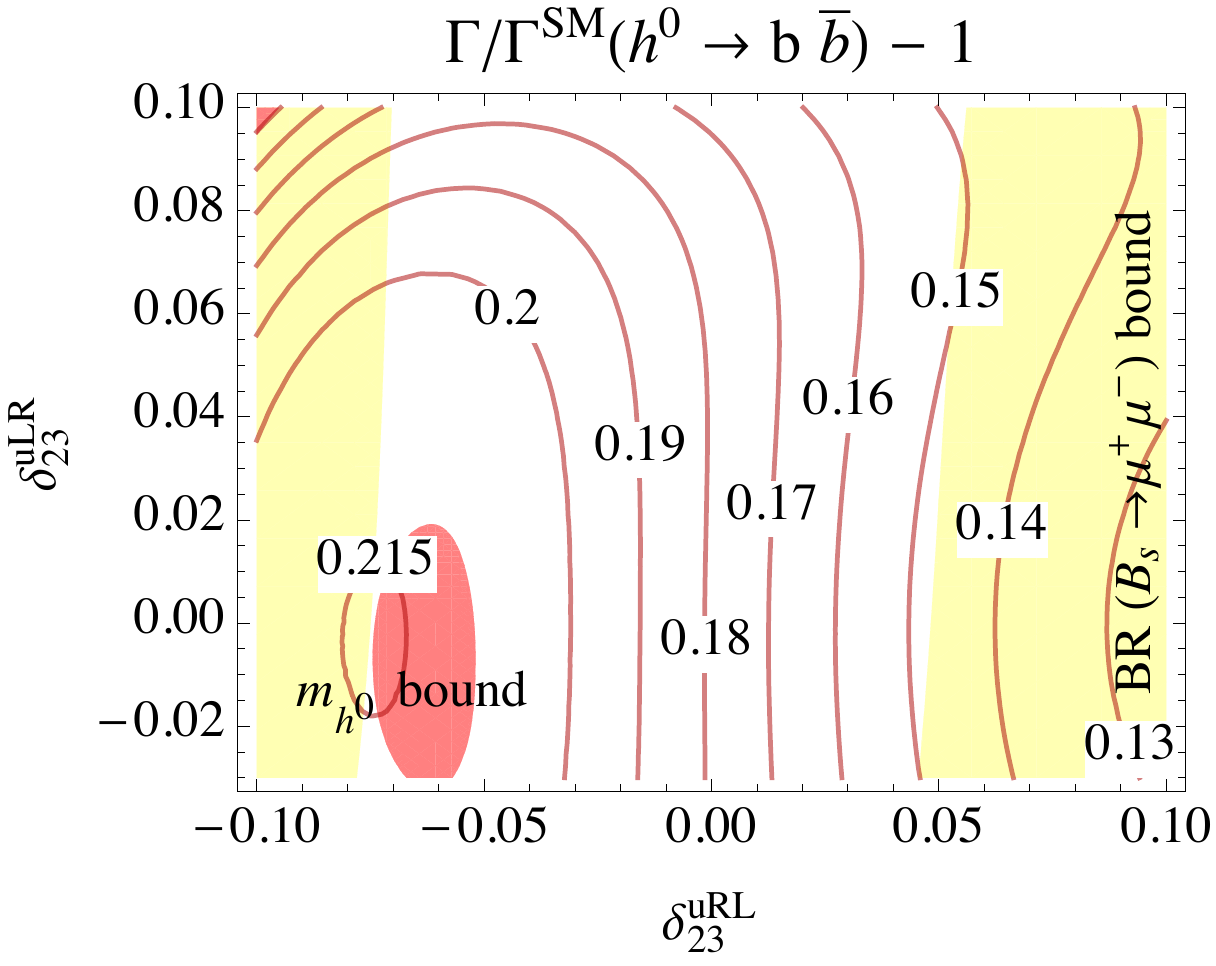}}}}
   \label{fig2a}
}
 \subfigure[]{
   { \mbox{\hspace*{+0.cm} \resizebox{7.cm}{!}{\includegraphics{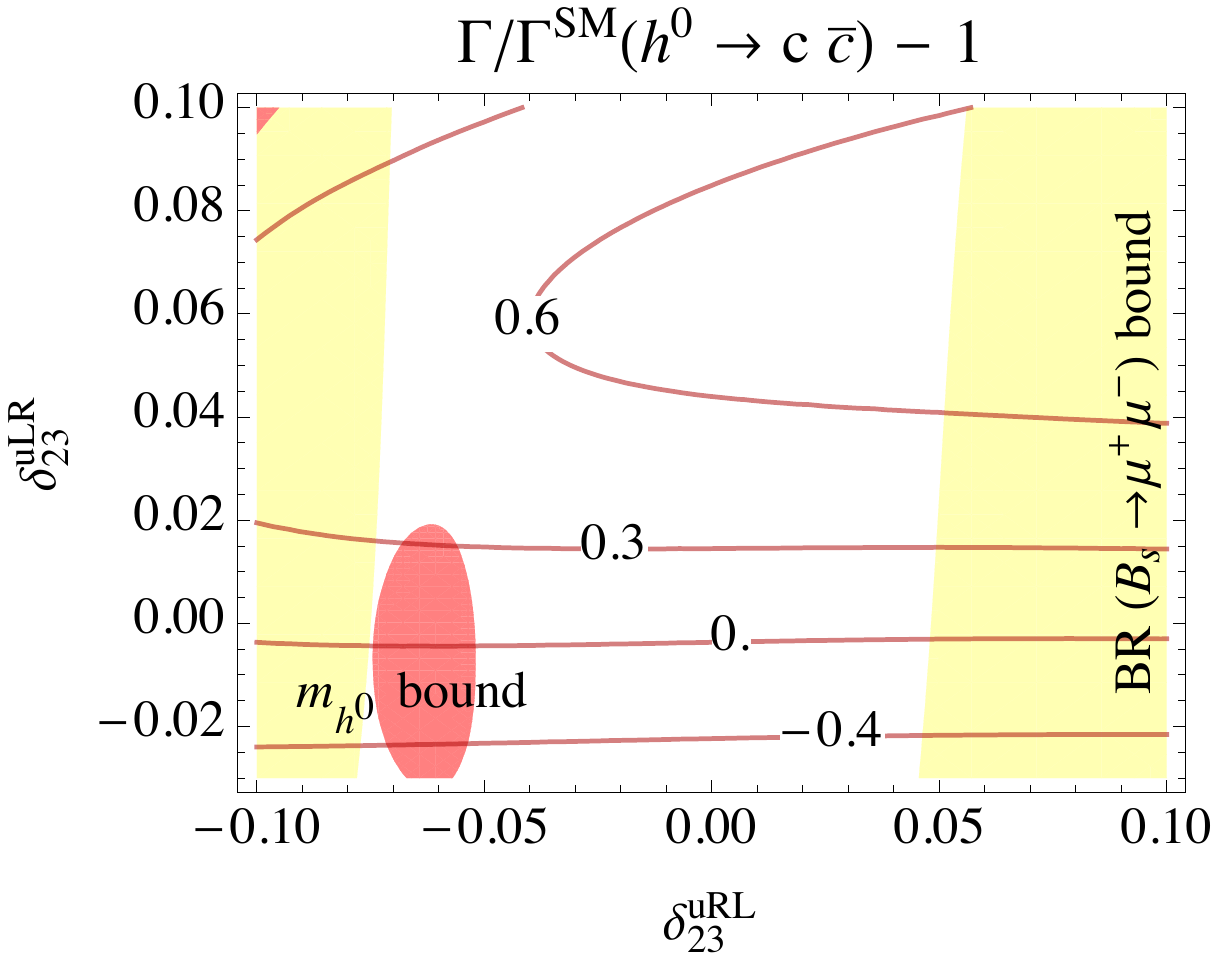}} \hspace*{-1cm}}}
  \label{fig2b}
}
\caption{Contours of the deviation (a)~$\Gamma/\Gamma^{\rm SM}(h^0 \to b \bar{b})-1$ and (b)~$\Gamma/\Gamma^{\rm SM}(h^0 \to c \bar{c})-1$ in the $\durl$-$\dulr$  plane for $\durr = 0.5$ and $\dll=0$.
}
\label{fig2}
\end{figure*}
\begin{figure*}[htb]
\centering
\subfigure[]{
   { \mbox{\hspace*{-1cm} \resizebox{7.cm}{!}{\includegraphics{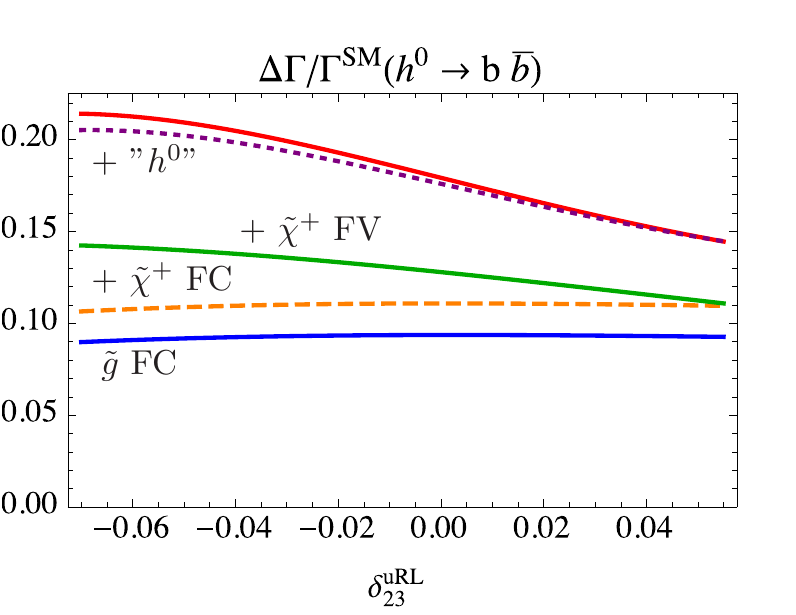}} }}
   \label{fig3a}
}
 \subfigure[]{
   { \mbox{\hspace*{+0.cm} \resizebox{7.cm}{!}{\includegraphics{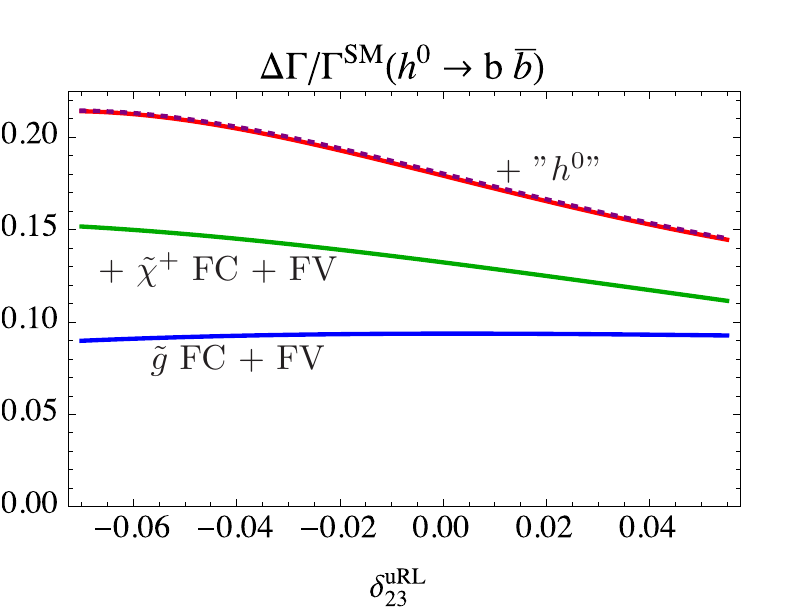}} \hspace*{-1cm}}}
  \label{fig3b}
  }
  \caption{Contours of the deviation (a)~$\Gamma/\Gamma^{\rm SM}(h^0 \to b \bar{b})-1$ in the $\durr$-$\durl$ plane for $\dll = 0$ and $\dulr=0$ and (b)~$\Gamma/\Gamma^{\rm SM}(h^0 \to c \bar{c})-1$ in the $\durr$-$\dulr$ plane for $\dll = 0$ and $\durl=0.02$.
}
\label{fig3}
\end{figure*}
Comparing the results for the charm self-energy, eqs.~(\ref{dsigLRsg_c1}), (\ref{MI_Ts}), and the vertex contribution to  $h^0 \to c \bar c$,
eqs.~(\ref{coeff_verth02ccb_MI}), (\ref{MI_Tvs}), we see that $T_2 = T^v_1 \hat v_2$. The same holds for the term proportional to $T^U_{33} M^Q_{23}  M^U_{23} $
in $T_3$ and $T^v_2$. Concerning the term proportional to $T^U_{33} T^U_{23}  T^U_{32}$, we have a factor 3 in the term  $T^v_2$ compared to that in $T_3$. Thus we can deduce the result $T^v_3$ from the term $T_4$ in eq.~(\ref{MI_Ts}) by adding a prefactor of 3 for all the terms with three $T^U$ elements. 

\subsection{Gluino and chargino contributions to $h^0 \to b \bar b$}
\label{SecMI:gluinocha2bbb}

Assuming the squared $\tilde d$-mass matrix $ {\cal M}^2_{\tilde{d}} \equiv M_{ij}$ in the form
\begin{equation}
\hspace*{-0.7cm}    \small{{\cal M}^2_{\tilde{d}} =          
         \left( \begin{array}{cccccc}
         M^{LL}_{11} & 0 & 0 & 0 & 0 & 0 \\
         0  & M^{LL}_{22} & M^Q_{23} & 0 & 0 & \hat v_1 T^D_{32} \\
         0 & M^Q_{23} & M^{LL}_{33} & 0 &  \hat v_1 T^D_{23} & M^{RL}_{33} \\
         0 & 0 & 0 &  M^{RR}_{11} & 0 & 0\\
         0 & 0 &   \hat v_1 T^D_{23} & 0 & M^{RR}_{22} &  M^D_{23} \\
         0 & \hat v_1 T^D_{32} & M^{RL}_{33}  & 0 &  M^D_{23} & M^{RR}_{33}
         \end{array}\right)},     
 \label{EqMassMatrix_sd}
\end{equation}
with $M^{RL}_{33}\sim - \mu m_b \tan\beta, \hat v_1 = v \cos\beta/\sqrt{2}$, and the QFV elements of the  $3 \times 3$ matrices $M^2_Q$ and $M^2_D$ are
denoted by $M^Q_{ij}$ and $M^D_{ij}$, respectively, for the $LR$-part of the gluino contribution to the bottom self energy $\Sigma^{LR, \tilde g}_b$, defined by the Lagrangian
${\cal L} = - \bar b\, \Sigma_b\, b$, we obtain
\begin{equation}
\hspace*{-0.7cm}
m_b \Sigma^{LR, \tilde g}_b =   - \frac{2 \a_s}{3 \pi } \msg (T^{FC}_1 + T^{FV}_2 + T^{FC}_3  + T^{FV}_3 + \ldots)
\label {dsigLRsg_b1}
\end{equation}
where the quark flavour conserving (FC) and quark flavour violating (FV) contributions read
\begin{eqnarray}
\hspace*{-1cm}
T^{FC}_1 & = & M^{RL}_{33}  \, b_0\! \left(1,  \msg^2, \{ M^{RR}_{33}, M^{LL}_{33}\} \right)\nn
T^{FV}_2 & = & \hat v_1 T^D_{32} M^Q_{23}  \, b_0\! \left(2,  \msg^2, \{ M^{RR}_{33}, M^{LL}_{22}, M^{LL}_{33}\} \right) \nn
              &+&   \hat v_1 T^D_{23} M^D_{23}  \, b_0\! \left(2,  \msg^2, \{M^{RR}_{33}, M^{RR}_{22}, M^{LL}_{33}\} \right) \nn
T^{FC}_3 & = & 3^\rho\, (M^{RL}_{33})^3  \nn
&& \times~b_0\! \left(3,  \msg^2, \{ M^{RR}_{33}, M^{LL}_{33}, M^{RR}_{33}, M^{LL}_{33}\} \right)\nn
T^{FV}_3 & = & (M^Q_{23})^2  M^{RL}_{33}   \nn
&&\times~ b_0\! \left(3,  \msg^2, \{ M^{RR}_{33}, M^{LL}_{33}, M^{LL}_{22}, M^{LL}_{33}\} \right)\nn
        & + &  (M^D_{23})^2  M^{RL}_{33}   \nn
        &&\times~ b_0\! \left(3,  \msg^2, \{ M^{RR}_{33}, M^{RR}_{33}, M^{RR}_{22}, M^{LL}_{33}\} \right)\nn
       & + & 3^\rho\, (\hat v_1)^2 (T^D_{32})^2  M^{RL}_{33} \nn
       &&\times~ b_0\! \left(3,  \msg^2, \{ M^{RR}_{33}, M^{LL}_{22}, M^{RR}_{33}, M^{LL}_{33}\} \right)\nn      
       & + & 3^\rho\, (\hat v_1)^2  (T^D_{23})^2  M^{RL}_{33} \nn
       &&\times~ b_0\! \left(3,  \msg^2, \{ M^{RR}_{33}, M^{LL}_{33}, M^{RR}_{22}, M^{LL}_{33}\} \right)
\end{eqnarray}     
with $\rho = 0$.
As in the $c\bar{c}$ case, the vertex contribution can be directly deduced from the self~energy,  with $T^{v\, x}_i = {T^{x}_i \over \hat v_1}, x = {FC, FV}$, 
an additional factor 3 for some terms in $T^x_3$,  $(M^{RL}_{33})^3  \to 3 (M^{RL}_{33})^3$, $(T^D_{23})^2 M^{RL}_{33} \to 3 (T^D_{23})^2 M^{RL}_{33}$,
$(T^D_{32})^2 M^{RL}_{33} \to 3 (T^D_{32})^2 M^{RL}_{33}$, and accordingly $\rho = 1$.

The relevant term for the self-energy calculation of the bottom-quark and for the vertex amplitude with a chargino in the loop
is proportional to $c^*_L c_R$, with $c_L = h_b U^*_{m 2} U^{\tilde u *}_{i 3}$ and
$c_R = - g V_{m 1} U^{\tilde u *}_{i 3} + h_t V_{m 2} U^{\tilde u *}_{i 6}$, where $U$ and $V$ diagonalize the chargino mass matrix $X$: $U^* X V^{-1} = M_D = {\rm diag}(m_{\tilde \chi^+_1}, m_{\tilde \chi^+_2})$. 
Neglecting the term
proportional to $g$ and $m_b$ in the loop integrals, we obtain
\bea
m_b \Sigma^{LR, \tilde \chi^+}_b &=&  {h_b h_t  \over 16 \pi^2} \sum_{m = 1}^2 \sum_{i = 1}^6  m_{\tilde \chi^+_m}  U_{m 2} V_{m 2}\, 
U^{\tilde u}_{i 3} U^{\tilde u *}_{i 6} \nn
&\times&~b_0(m^2_{ \tilde \chi^+_m}, m^2_{\tilde u_i})\, .
\label{mbchaMI1}
\eea
The mass insertions in the $\tilde u_i$ line can be deduced from the results for the bottom self-energy with gluino in the loop. Moreover, we can also apply the MI technique to the chargino part ($\sim  U_{m 2} V_{m 2}$) in eq.~(\ref{mbchaMI1}) using linear approximation.  
Finally, we obtain the approximate
result
\begin{equation}
\hspace*{-0.7cm}
m_b \Sigma^{LR, \tilde \chi^+}_b =  {h_b h_t  \over 16 \pi^2} \mu\,  
\left(T^{FC}_1 + T^{FV}_2 + T^{FC}_3  + T^{FV}_3 + \ldots\right),
\label{mbchaMI2}
\end{equation} 
where the explicit result for the terms $T^x_{i}$ (see Ref.~\cite{Eberl:2016aox}) is lengthy and therefore not shown here.
\section{Numerical analysis}
\label{sec:num}

To demonstrate the effects of QFV we have chosen a reference scenario with strong $\ti{c}-\ti{t}$ mixing in both $h^0 \to b\bar{b}$ and $ h^0\to c\bar{c}$ decays.
The corresponding MSSM parameters at $Q = 1$~TeV are shown in Table~\ref{basicparam}. 
%
\begin{table}
\caption{Reference scenario: shown are the basic MSSM parameters 
at $Q = 1$~TeV,
except for $m_{A^0}$ which is the pole mass of $A^0$, 
with $T_{U33} = 1450$~GeV (corresponding to $\delta^{uRL}_{33} =  0.1$). All other squark 
parameters not shown here are zero. }
\begin{center}
\small{
\begin{tabular}{|c|c|c|}
  \hline
 $M_1$ & $M_2$ & $M_3$ \\
 \hline \hline
400 ~\gev  & 800 ~\gev & 2000~\gev \\
  \hline
\end{tabular}
\vskip0.4cm
\begin{tabular}{|c|c|c|}
  \hline
 $\mu$ & $\tan \beta$ & $m_{A^0}$ \\
 \hline \hline
 500~\gev & 30  &  1500 ~\gev \\
  \hline
\end{tabular}
\vskip0.4cm
\begin{tabular}{|c|c|c|c|}
  \hline
   & $\alpha = 1$ & $\alpha= 2$ & $\alpha = 3$ \\
  \hline \hline
   $M_{Q \alpha \alpha}^2$ & $3200^2~\gev^2$ &  $1550^2~\gev^2$  & $1100^2~\gev^2$ \\
   \hline
   $M_{U \alpha \alpha}^2$ & $3200^2~\gev^2$ & $2800^2~\gev^2$ & $2050^2~\gev^2$ \\
   \hline
   $M_{D \alpha \alpha}^2$ & $3200^2~\gev^2$ & $3000^2~\gev^2$ &  $2500^2~\gev^2$  \\
   \hline
\end{tabular}
\vskip0.4cm
\begin{tabular}{|c|c|c|c|}
  \hline
   $\delta^{LL}_{23}$ & $\delta^{uRR}_{23}$  &  $\delta^{uRL}_{23}$ & $\delta^{uLR}_{23}$\\
  \hline \hline
  0  & 0.8 & 0.02  & 0.02  \\
    \hline
\end{tabular}}
\end{center}
\label{basicparam}
\end{table}
%
%
\begin{table}
\caption{Physical masses in GeV of the particles for the scenario of Table~\ref{basicparam}.}
\begin{center}
\small{
\begin{tabular}{|c|c|c|c|c|c|}
  \hline
  $\mnt{1}$ & $\mnt{2}$ & $\mnt{3}$ & $\mnt{4}$ & $\mch{1}$ & $\mch{2}$ \\
  \hline \hline
  $395$ & $507$ & $511$ & $845$ & $501$ & $845$ \\
  \hline
\end{tabular}
\vskip 0.4cm
\begin{tabular}{|c|c|c|c|c|}
  \hline
  $m_{h^0}$ & $m_{H^0}$ & $m_{A^0}$ & $m_{H^+}$ \\
  \hline \hline
  $125$  & $1500$ & $1500$ & $1503$ \\
  \hline
\end{tabular}
\vskip 0.4cm
\begin{tabular}{|c|c|c|c|c|c|c|}
  \hline
  $\msg$ & $\msu{1}$ & $\msu{2}$ & $\msu{3}$ & $\msu{4}$ & $\msu{5}$ & $\msu{6}$ \\
  \hline \hline
  $2103$ & $996$ & $1176$ & $1578$ & $3214$ & $3217$ & $3327$ \\
  \hline
\end{tabular}
\vskip 0.4cm
\begin{tabular}{|c|c|c|c|c|c|}
  \hline
 $\msd{1}$ & $\msd{2}$ & $\msd{3}$ & $\msd{4}$ & $\msd{5}$ & $\msd{6}$ \\
  \hline \hline
  $1128$ & $1579$ & $2515$ & $3012$ & $3211$ & $3218$ \\
  \hline
\end{tabular}}
\end{center}
\label{physmasses}
\end{table}
%
\begin{table}
\caption{Flavour decomposition of $\su_i$, $i=1,...,6$ for the scenario of Table~\ref{basicparam}. Shown are the squared coefficients. }
\begin{center}
\small{
\begin{tabular}{|c|c|c|c|c|c|c|c|}
  \hline
  & $\su_L$ & $\sca_L$ & $\st_L$ & $\su_R$ & $\sca_R$ & $\st_R$ \\
  \hline \hline
 $\su_1$ & $0$ & $0.002$ & $0.25$ & $0$ & $0.228$ & $0.52$  \\
  \hline 
  $\su_2$  & $0$ & $0$ & $0.749$ & $0$ & $0.086$ & $0.165$ \\
  \hline
  $\su_3$  & $0.051$ & $0.946$ & $0.001$ & $0$ & $0$ & $0$ \\
  \hline
  $\su_4$ & $0.95$ & $0.05$ & $0$ & $0$ & $0$ & $0$  \\
  \hline 
  $\su_5$  & $0$ & $0$ & $0$ & $1$ & $0$ & $0$ \\
  \hline
  $\su_6$  & $0$ & $0$ & $0$ & $0$ & $0.69$ & $0.31$ \\
  \hline
\end{tabular}}
\end{center}
\label{flavourdecomp}
\end{table}
This scenario satisfies all present experimental and theoretical constraints explicitely listed in Ref.~\cite{Eberl:2016aox}. The resulting physical masses of the particles are shown in Table~\ref{physmasses}. The flavour decomposition of the up-type squarks $\su_i, i=1,...,6$ is shown in Table~\ref{flavourdecomp}. In the following, unless specified otherwise, we show various parameter dependences of the relative to the SM width $\Gamma/\Gamma^{\rm SM}-1$ for $\Gamma(h^0 \to b \bar{b})$ and $\Gamma(h^0 \to c \bar{c})$ with all other parameters fixed as in Table~\ref{basicparam}. 

In Fig.~\ref{fig2} the dependence on the QFV parameters $\durl$ and $\dulr$ is shown. It is seen that in the case of $b\bar{b}$ (Fig.~\ref{fig2a}) the variation due to correlated $\sca_R-\st_L$ and $\sca_L-\st_R$ mixing can vary up to $\sim 7\%$ in the region allowed by the constraints. Comparing Fig.~\ref{fig2a} with Fig.~\ref{fig2b} one can see that there exist regions where both widths considered simultaneously deviate significantly from their SM prediction. Hence $\Gamma(h^0 \to b \bar{b})$ tends to depend more on $\sca_R-\st_L$ mixing, while $\Gamma(h^0 \to c \bar{c})$ depends more on $\sca_L-\st_R$ mixing. 

In Section~\ref{SecMI:gluino2ccb}, in agreement with our results in Ref.~\cite{Bartl:2014bka}, we have shown that in the case of $c\bar{c}$ the deviation from the SM is entirely due to QFV. However, it is known that in the MSSM $\Gamma(h^0 \to b\bar{b})$ can differ considerably from the SM due to quark-flavour conserving (QFC) contributions~\cite{Endo:2015oia}. 
In Fig.\ref{fig3}(a) the individual one-loop contributions to $\Gamma(h^0 \to b\bar{b})$ as a function of $\durl$ are shown. The top curve shows the full one-loop contribution to the width with no approximation. It is seen that the main one-loop contribution to $\Gamma(h^0 \to b\bar{b})$ comes from QFC gluino and chargino exchange. Nevertheless, there exist a region for large and negative $\durl$ where the QFV component can be comparable with the QFC component. The QFV component is mainly due to chargino exchange which involves mixing in the $\su$-sector. On the other hand, the QFV gluino exchange, which plays a major role in the $c \bar{c}$ case, in the $b \bar{b}$ case involves $\sd$ quarks whose mixing is strongly suppressed, and hence, this contribution is very small and thus not considered here. 
It is also interesting that the QFV component receives a large contribution from the dependence of 
$\Gamma^{g, \rm impr}$ on the Higgs mass and the angle $\alpha$, which depend on the QFV parameters. 
In Figs.~\ref{fig3}(a) and ~\ref{fig3}(b) this contribution is denoted by $"h^0"$.
Note that $m_{h^0}$ as well as $\sin \alpha$ already appear in the kinematics factor at tree level, see eq.~(\ref{decaywidttree}). The total QFV contribution to  $\Gamma/\Gamma^{\rm SM}(h^0 \to b\bar{b})$ can be as large as $\sim 8\%$ at a certain point. Fig.\ref{fig3}(b), where no MI is used, demonstrates in addition the quality of the total approximated result obtained in~\cite{Eberl:2016aox}, see eq.~(4.38) therein.
A numerical comparison of the different MI orders shows that the MI formulas converge
fast for $\tilde g$~FC and $\tilde \chi^+$~FC, but not for $\tilde \chi^+$~FV, compare Fig.~\ref{fig3}(a) with Fig.~\ref{fig3}(b).
The difference between the dotted curve and the
upper curve in Fig.~\ref{fig3}(a), therefore, is mainly due to the relatively slow MI convergence of the $\tilde \chi^+$~FV contribution.

Although the decay $h^0 \to b \bar{b}$ is dominant, the measurement of its branching ratio and width at the LHC will be hard due to the huge QCD background. In any case, high luminosity at LHC would be needed~\cite{CMS:2013xfa}. A model independent and precise measurement of B($h^0 \to b \bar{b}$) and $\Gamma(h^0 \to b \bar{b})$ would be possible at a $e^+ e^-$ linear collider such as ILC~\cite{Barklow:2015tja}. 

\section{Conclusions}
\label{sec:num}

We have studied the decays $h^0 \to b\bar{b}$ and $h^0 \to c\bar{c}$ at full one-loop level in the MSSM with quark-flavour mixing in the heavy squark sector. The dominant contributions with gluino and chargino exchange are calculated in the mass insertion approximation, using the Flavour Expansion Theorem. Both widths, $\Gamma(h^0 \to b \bar{b})$ and $\Gamma(h^0 \to c \bar{c})$, can deviate from the SM significantly within the allowed parameter region. In the $c\bar{c}$ case the deviation is mainly due to the MSSM QFV parameters. In the $b\bar{b}$ case the deviation is mainly due to the MSSM QFC parameters, but nevertheless at certain parameter regions the the QFV parameters can cause fluctuations of $\Gamma(h^0 \to b \bar{b})$ up to $\sim 7\%$, with similar large contribution coming from the Higgs parameters dependence. 





\begin{thebibliography}{00}
 
\bibitem{Eberl:2016aox}
  H.~Eberl, E.~Ginina, A.~Bartl, K.~Hidaka and W.~Majerotto,
  JHEP 1606 (2016) 143
  [arXiv:1604.02366 [hep-ph]].

\bibitem{Allanach:2008qq}
  B.~C.~Allanach {\it et al.},
  Comput. Phys. Commun. 180 (2009) 8
  [arXiv:0801.0045 [hep-ph]].
  
\bibitem{Bartl:2014bka}
  A.~Bartl, H.~Eberl, E.~Ginina, K.~Hidaka and W.~Majerotto,
  Phys. Rev. D 91 (2015) no.1,  015007
  [arXiv:1411.2840 [hep-ph]].

\bibitem{Dedes:2015twa}
  A.~Dedes, M.~Paraskevas, J.~Rosiek, K.~Suxho and K.~Tamvakis,
  JHEP 1506 (2015) 151
  [arXiv:1504.00960 [hep-ph]].
  
\bibitem{Endo:2015oia}
  M.~Endo, T.~Moroi and M.~M.~Nojiri,
  JHEP 1504 (2015) 176
  [arXiv:1502.03959 [hep-ph]].
  
\bibitem{CMS:2013xfa}
  [CMS Collaboration],
  [arXiv:1307.7135 [hep-ex]].

 
\bibitem{Barklow:2015tja}
  T.~Barklow, J.~Brau, K.~Fujii, J.~Gao, J.~List, N.~Walker and K.~Yokoya,
  [arXiv:1506.07830 [hep-ex]].





\end{thebibliography}



\end{document}